\documentclass[11pt]{article}

\usepackage{graphicx}
\usepackage{epstopdf}
\usepackage{amsmath}
\usepackage{amssymb}
\usepackage{amsfonts}
\usepackage{amsthm}
\usepackage[usenames]{color}
\usepackage{array}

\usepackage[
      colorlinks=true,
      linkcolor=blue,
      urlcolor=blue,
      filecolor=blue,
      citecolor=red,
      pdfstartview=FitV,
      pdftitle={},
      pdfauthor={},
      pdfsubject={},
      pdfkeywords={},
      pdfpagemode=None,
      bookmarksopen=true
]{hyperref}

\usepackage{epsfig}

\usepackage{hyperref}

\newcommand{\be}{\begin{equation}}
 \newcommand{\ee}{\end{equation}}
 \newcommand{\bea}{\begin{eqnarray}}
 \newcommand{\eea}{\end{eqnarray}}

\renewcommand{\(}{\left(}
\renewcommand{\)}{\right)}

\numberwithin{equation}{section}

\begin{document}

\begin{flushright}
\texttt{\today}
\end{flushright}

\begin{centering}

\vspace{2cm}

\textbf{\Large{
Aspects of Ultra-Relativistic Field Theories\hspace{5cm} via Flat-space Holography
%and Flat Space Holography
}}
 \vspace{0.8cm}
 
  {\large  Reza Fareghbal$^a$, Ali Naseh$^b$, Shahin Rouhani$^{c,b}$ }

  \vspace{0.5cm}

\begin{minipage}{.9\textwidth}\small
\begin{center}
{\it $^a$ Department of Physics, 
Shahid Beheshti University, 
G.C., Evin, Tehran 19839, Iran.  }\\
{\it $^b$ School of Particles and Accelerators, Institute for Research in Fundamental Sciences (IPM)\\
P.O. Box 19395-5531, Tehran, Iran }\\
{\it ${}^c$ Department of Physics, Sharif University of Technology, P.O. Box 11365-9161, Tehran, Iran}\\
  \vspace{0.5cm}
{\tt r$\_$fareghbal@sbu.ac.ir, naseh@ipm.ir, rouhani@ipm.ir}
\\ $ \, $ \\
\end{center}
\end{minipage}
%\end{center}
\begin{abstract}
Recently it was proposed that asymptotically flat spacetimes have a holographic dual which is an ultra-relativistic conformal field theory. In this paper, we obtain the conformal anomaly for such a theory via the flat-space holography technique. Furthermore, using flat-space holography we obtain a $C$-\hspace{.5mm}function for this theory which is monotonically decreasing from the UV to the IR by employing the null energy condition in the bulk. 
	
\end{abstract}

\end{centering}

\newpage

\tableofcontents

%\newpage

%\setcounter{equation}{0}

\section{Introduction}
It is an interesting question to extent the celebrated AdS/CFT correspondence \cite{Maldacena:1997re} to the spacetimes other than AdS. One possible direction is exploration of holography for the asymptotically flat spacetimes. Since these spacetimes are given by taking zero cosmological constant limit from the asymptotically AdS spacetimes, it is of interest to look for the corresponding operation of flat-space limit in the boundary theory. It is proposed in papers \cite{{Bagchi:2010zz}} and \cite{{Bagchi:2012cy}} that the flat-space limit (zero cosmological constant limit) at the bulk side corresponds to taking ultra-relativistic limit of a dual CFT. The key point in this correspondence is the equivalence between symmetries of ultra-relativistic theory and asymptotic symmetries of asymptotically flat spacetimes at null infinity. These symmetries which are known as Bondi-Metzner-Sachs (BMS) symmetries \cite{BMS} are infinite dimensional at three \cite{Ashtekar:1996cd,Barnich:2006av} and four dimensions \cite{Barnich:4dBMS}.

The Flat/Ultra-relativistically contracted CFT correspondence can be used to get insight about the quantum nature of gravity in the asymptotically flat spacetimes or answer some questions about the ultra-relativistic theories by using the holographic calculations (see recent papers \cite{Bagchi:2015wna,Hosseini:2015uba} and references therein ). In this paper we want to use this duality and study some aspects of the ultra-relativistic theories.
In particular we will zoom in on 2d ultra-relativistic field theories. This symmetry arises as a contraction of the Poincare symmetry, in the limit of vanishing velocity of light.

 A related  question in these field theories  is to find a quantity which monotonically decreases with renormalization group (RG) flow. The manifestation of scale dependence in quantum field theory is the renormalization group (RG) flow of coupling constants with scale. It is however not clear why this irreversible flow (the name group should better read semi-group) should flow into a fixed point, while in principal, strange attractors are possible. That these pathological cases do not happen in physically relevant systems is interesting. In 2d this is due to Zamolodchikov's c-theorem \cite{Zamolodchikov:1986gt}, and in 4d due to a-theorem\cite{Luty:2012ww,Komargodski:2011vj}\footnote{Further discussions can be found in \cite{Dymarsky:2013pqa}-\cite{Naseh:2016maw}.}. In other words the existence of a decreasing function on the RG flow enforces flow towards a fixed point (or limit cycle). There is an intuitive way of understanding the IR fixed point. In Wilsons renormalization group viewpoint \cite{Wilson:1974mb}, at any non-trivial energy scale, we simply integrate out the fast degrees of freedom. Eventually, we should not have any degrees of freedom left, in other words degrees of freedom reduce with scale, which is the essence of Cardy's formula \cite{Cardy:1986ie}. 

It may be argued that for ultra-relativistic field theories, even if existence of fixed points cannot be proven, one is on safe grounds since in some limit the relativistic theory is approached and RG flow would be well behaved. However one expects that the existence of a c-theorem should also be provable in ultra-relativistic field theories. In this paper, using connection between ultra-relativistic field theory and asymptotically flat spacetimes, we propose a $C$-\hspace{.5mm}function which  is monotonically decreasing from the UV to the IR. Our starting point is  AdS/CFT and holographic calculation of $C$-\hspace{.5mm}function for a theory which is dual of asymptotically AdS spacetimes. We take the flat-space limit in this calculation and use the null energy condition in the bulk to find the $C$-\hspace{.5mm}function of the ultra-relativistic theory. We also use holographic method and calculate the trace anomaly of ultra-relativistic theory. The interesting point is that similar to the relativistic CFT, the conformal anomaly is related to the curvature of the spacetimes on which ultra-relativistic theory lives. 

In section two we start with preliminaries and try to clarify the connection between ultra-relativistic contraction of CFT$_{2}$ and flat space limit of AdS$_{3}$ spacetimes. Section three is devoted to holographic calculations where we find conformal anomaly of the ultra-relativistic  theory. In section four we  propose a $C$-\hspace{.5mm}function for it.
\section{Ultra-relativistic CFT and flat-space holography}
In this paper we are interested in ultra-relativistic contraction of 2d CFTs, which is achievable if speed of light tends to zero.   Another equivalent way of taking the ultra-relativistic limit of CFT is performed by scaling time; $t\to\epsilon t$ and taking $\epsilon\to 0$  limit \cite{Bagchi:2012cy}. If one starts with the generators of $SO(2,2)$ and contract it by scaling of time, the final algebra is given by \cite{Bagchi:2012cy}
\begin{eqnarray}\label{carroll-finite}
&& [E,P] = 0,~~~ [E,D] = E ,~~~ [E,B] = 0,~~~ [E,F] = 0,~~~ [E,G] = -2B ,\cr \nonumber \\&&
[P,D] = P,~~~ [P,B] = -E,~~~ [P,F] = -2B ,~~~ [P,G] =2D,~~~[D,B] =0,\cr\nonumber\\ &&
~~~ [D,F] = F,~~~ [D,G] = G,~~~ [B,F] =0,~~~ [B,G] =-F,~~~ [F,G] =0.
\end{eqnarray}
 The generators $E,P,B,D,F,G$ are respectively  energy, momentum,  boost,  dilation,  acceleration and special conformal transformation (for their precise definitions  see below (\ref{D1})).  Note  that this algebra is isomorphic to Galilean Conformal Algebra (GCA) \cite{Bagchi:2009my} simply by switching the space and time coordinates. A duality which seems to be peculiar to 1+1 dimensions. The above algebra \eqref{carroll-finite}  has an affine extension with an infinite number of generators which is given by contraction of two copies of Virasoro algebra. More precisely, if we start with two copies of the conformal algebra with generators $l_{n},\bar{l}_{n}$ and central charges $c,\bar{c}$ and define generators \cite{Bagchi:2012cy,Bagchi:2009my,Hosseiny:2009jj}
\begin{eqnarray}\label{def.of.contraction}
M_{n} &=& \lim_{\epsilon\rightarrow 0} \epsilon\left(l_{n}+\bar{l}_{-n}\right),\cr\nonumber
L_{n} &=& \lim_{\epsilon\rightarrow 0} \left(l_{n}-\bar{l}_{-n}\right),
\end{eqnarray} 
the final algebra is 
\begin{align}\label{2d carroll algebra}
\nonumber [L_m,L_n]&=(m-n)L_{m+n}+c_L m(m^2-1)\delta_{m+n,0}, \\ 
  [L_m,M_n] &= (m-n)M_{m+n}+c_M m(m^2-1)\delta_{m+n,0},  
  \end{align}
  where
  \begin{eqnarray}\label{non.rel.cent.charge}
c_L &=& \lim_{\epsilon\rightarrow 0} \frac{1}{12} (c-\bar{c}),\cr\nonumber
c_{M} &=& \lim_{\epsilon\rightarrow 0} \frac{\epsilon}{12} (c+\bar{c}).
\end{eqnarray}
The generators of the global part of the algebra \eqref{2d carroll algebra} are related to \eqref{carroll-finite} by
\begin{align}\label{D1}
L_0=D,\qquad L_1=G,\qquad L_{-1}=-P,\nonumber\\
M_0=-B,\qquad M_1=F,\qquad M_{-1}=E.
\end{align}

The interesting point is that \eqref{2d carroll algebra} appears also  in a completely different context. It is the asymptotic symmetry at the null infinity of the  three dimensional  asymptotically   flat spacetimes \cite{Barnich:2006av}. Because of this, it was proposed  in \cite{Bagchi:2010zz}-\cite{Bagchi:2012cy} that the holographic dual of asymptotically flat spacetimes is a field theory which has ultra-relativistic conformal symmetry. The dual ultra-relativistic  field theory is given by contraction of CFT. Since according to AdS/CFT correspondence the parent CFT is dual to asymptotically AdS spacetimes, the contracted CFT corresponds to flat-space limit (zero cosmological constant limit) of the asymptotically AdS spacetimes. If one starts with AdS$_3$ spacetime given by 
\begin{equation}\label{conformal factor.1}
  ds^2=-\(1+{r^2\over\ell^2}\)d\tau^2+{dr^2\over\(1+{r^2\over\ell^2}\)}+r^2d\phi^2,
  \end{equation}  
 the geometry which  the dual CFT lives on it, is given by using  conformal boundary of the AdS spacetime,
 \begin{equation}\label{conformal factor}
  ds^2_{boundary}= {r^2\over G^2}\(-{G^2\over\ell^2}d\tau^2+G^2d\phi^2\)
  \end{equation} 
where we have intentionally used Newton's constant in the conformal factor to make it dimensionless and also have a $\ell$ independent conformal factor. It is clear from \eqref{conformal factor} that the boundary CFT lives on a two dimensional flat spacetimes and its  time coordinate, $t$, is related to the bulk as $t={G\over\ell}\tau$. Now taking flat-space limit in the bulk side by sending $\ell\to\infty$ or $\epsilon={G\over\ell}\to 0$ corresponds to contraction of time in the boundary as $t\to\epsilon t$.
%\footnote{The relation $t={G\over\ell}\tau$ between the time coordinates of the bulk and the boundary theories imposes $c_{\text{boundary}}={G\over\ell} c_{\text{bulk}}$ where $c_{\text{bulk}}$ is  speed of light in the bulk and $c_{\text{boundary}}$ is speed of light at the boundary which is measured from the point of view of the bulk observer. It is clear that by taking  flat limit ${G\over\ell}\to 0$, $c_{\text{boundary}}$ tends to zero.  }.
 Using this correspondence one can study some aspects of gravity in the three dimensional asymptotically flat spacetimes by using two dimensional  ultra-relativistically contracted CFT. For example a quasi local stress tensor for asymptotically flat spacetimes is achievable  if we start with energy momentum of a CFT and contract it \cite{Fareghbal:2013ifa}. 

To be more precise , let us consider  energy momentum tensor of a  relativistic CFT, $T_{\mu\nu}$,  which satisfies a relativistic conservation equation . The components of the energy momentum tensor of  the contracted CFT, $\tilde T_{\mu\nu}$ are given by \cite{Fareghbal:2013ifa}
\begin{align}\label{def.of contracted EM}
\nonumber\tilde T_{++}+\tilde T_{--}&=\lim_{\epsilon\to 0}{\epsilon}\left(T_{++}+ T_{--}\right)\\
\nonumber\tilde T_{++}-\tilde T_{--}&=\lim_{\epsilon\to 0}\left(T_{++}- T_{--}\right)\\
\tilde T_{+-}&=\lim_{\epsilon\to 0}{\epsilon\, T_{+-}}
\end{align} 
where $+$ and $-$ correspond to light-cone coordinates in both of the theories. 
 
\section{Ultra-relativistic  conformal anomaly}
 In this section we calculate the conformal anomaly of the ultra-relativistic conformal field theories using holography. The starting point for the  holographic calculations is AdS/CFT and its standard dictionary for the relation between bulk and boundary quantities and then taking the flat-space limit.  However, the flat-space limit or taking  infinite radius limit of asymptotically AdS spacetimes is gauge dependent and is not well-defined for line elements written in the Fefferman-Graham coordinates. In order to have a well-defined flat limit all calculations will be done in the so called BMS gauge \cite{Barnich:2012aw}. For the three-dimensional asymptotically locally AdS spacetimes  the BMS gauge  is given by,
\begin{equation}\label{ALAdS ansatz}
    ds^2=\mathcal{A}(u,r,\phi)du^2-2 e^{2\beta(u,\phi)}dudr+2\mathcal{B}(u,r,\phi)du d\phi+r^2d\phi^2.
    \end{equation}   
where for the  Einstein gravity with negative cosmological constant,
\begin{equation}\label{action}
 S={1\over 16\pi G}\int\, d^3x \sqrt{-g}\left(R+{2\over\ell^2}\right),
 \end{equation} 
 $\mathcal{A}$ and $\mathcal{B}$ are determined as
\begin{align}\label{solution.ads}
\nonumber\mathcal{A}(u,r,\phi)&= -e^{4\beta(u,\phi)} {r^2\over \ell^2}+\mathcal{M}(u,\phi),\\
\mathcal{B}(u,r,\phi)&= -r\partial_\phi e^{2\beta(u,\phi)}+\mathcal{N}(u,\phi),
\end{align}
and
\begin{align}\label{eom.ads}
\nonumber\partial_\phi\mathcal{M}&-2\partial_u\mathcal{N}+4\mathcal{N}\partial_u\beta=0\\
-8\partial_\phi\beta\partial_u\partial_\phi\beta&-4\partial_u\partial_\phi^2\beta+{4\over\ell^2}\mathcal{N}\partial_\phi\beta+{2\over\ell^2}\partial_\phi\mathcal{N}-e^{-4\beta}\partial_u\mathcal{M}+4\mathcal{M}e^{-4\beta}\partial_u\beta=0.
\end{align}

The quasi-local stress tensor $T_{\mu\nu}$ of the metric \eqref{ALAdS ansatz} are given by using the standard calculation of the holographic renormalization method \cite{Balasubramanian:1999re,HH}. According to AdS/CFT correspondence the components of $T_{\mu\nu}$ at the bulk correspond to expectation values of the energy- momentum tensor of the boundary CFT. These components are 
\begin{align}\label{EM.tensor.ads}
\nonumber T_{uu}&=\frac{e^{4\beta}\left(\partial_\phi^2\beta+(\partial_\phi\beta)^2\right)}{4\pi\ell G}+\frac{{\mathcal{M}}}{16\pi\ell G},\\
\nonumber T_{u\phi}&={\mathcal{N}\over 8\pi\ell G},\\
T_{\phi\phi}&=\frac{\ell\mathcal{M}e^{-4\beta}}{16\pi G}-\frac{\ell (\partial_\phi\beta)^2 }{4\pi G}.
\end{align}
 
 Conformal boundary of the asymptotically locally AdS spacetimes \eqref{ALAdS ansatz} is given by
 \begin{equation}\label{CB}
 ds^2|_{C.B}=-\frac{G^2}{\ell^2}e^{4\beta(u,\phi)}du^2+G^2d\phi^2
 \end{equation}
where we have intentionally used $G$ to make the conformal factor (which was used in \eqref{ALAdS ansatz} for defining conformal boundary) dimensionless. It is clear from \eqref{CB} that the conformal boundary is non-flat. Thus the dual CFT must live on a non-flat spacetime. The Ricci scalar of the conformal boundary is
\begin{equation}\label{ricci.ads}
   R_{C.B}=-{4\over G^2}\left(\partial_\phi^2\beta+2(\partial_{\phi}\beta)^2\right).
   \end{equation}   
Using \eqref{EM.tensor.ads} and \eqref{ricci.ads} we can calculate the trace of stress tensor as
\begin{equation}
T=g^{ij}T_{ij}=-\frac{\ell}{4\pi G^3}\left(\partial_\phi^2\beta+2(\partial_{\phi}\beta)^2\right),
\end{equation}
which can be  written as
\begin{equation}\label{conformal.anomaly}
T=\dfrac{C}{24\pi}R_{C.B},
\end{equation}
where $C=3\ell/2G$ is the Brown and Henneaux's central charge. The result  \eqref{conformal.anomaly}  is the conformal anomaly of the boundary CFT. 

We should note that the $u$ dependence in function $\beta(u,\phi)$ is crucial to have  conformal anomaly. For the case which $\beta$ is only a function of $\phi$ coordinate we have
%\bea
%R_{C.B} = -4\Box \beta(\phi),
%\eea
%and therefore
\bea\label{IM}
T = \Box \Phi(\phi),
\eea
where $\Box \equiv \nabla^{i}\nabla_{i}$  is defined by the metric (\ref{CB}) and $\Phi(\phi)
= -\dfrac{C}{6\pi}\beta(\phi)$. Appearance of $\Box \Phi(\phi)$ on the right-hand side
of (\ref{IM}) creates the opportunity to define an improved stress tensor:
\bea
\Theta_{ij} = T_{ij} + \left(\nabla_{i}\nabla_{j}
-g_{ij}\Box\right)\Phi(\phi),
\eea
which is traceless. 
%It is easy to see that this improved stress tensor is not conserved. Apart
%from the problem of non-conservation, we have shown that if a function $\beta$ depends just
%on the $\phi$ coordinate, the non-zero term in the right hand side of (\ref{IM}) is
%artificial and the theory is a CFT. 
Therefore to have a non-trivial
conformal anomaly
%\footnote{A non-trivial conformal anomaly 
%is one which can not be removed by adding improvement terms to the stress %tensor.} 
(\ref{conformal.anomaly}), "$u$" dependence of the function $\beta$ is crucial.

Now take flat-space limit in the above calculations and propose a holographic derivation for the ultra-relativistic conformal anomaly. The main steps have been done in paper \cite{Fareghbal:2013ifa} where the stress tensor near the null infinity of  the asymptotically flat spacetimes $\tilde T_{ij}$ is related to the AdS counterpart $T_{ij}$ as
\begin{align}\label{def.of.flat EM}
\nonumber\tilde T_{++}+\tilde T_{--}&=\lim_{{G\over\ell}\to 0}{G\over\ell}\left(T_{++}+ T_{--}\right)\\
\nonumber\tilde T_{++}-\tilde T_{--}&=\lim_{{G\over\ell}\to 0}\left(T_{++}- T_{--}\right)\\
\tilde T_{+-}&=\lim_{{G\over\ell}\to 0}{G\over\ell}T_{+-}
\end{align}
where $x^\pm={u \over \ell}\pm\phi$ for the asymptotically AdS solutions and $x^\pm={u\over G}\pm\phi$ for the asymptotically flat cases. We should note that in the BMS gauge, the asymptotically flat solutions are given also as \eqref{ALAdS ansatz}. However,  $\mathcal{A}$ and $\mathcal{B}$ are given by taking flat limit i.e. $G/\ell\to 0$, from the equations  \eqref{solution.ads} and \eqref{eom.ads}. Another point is that according to \cite{Fareghbal:2013ifa} the dual ultra-relativistic theory lives on a two dimensional spacetime which is given by taking flat limit from the conformal boundary of the AdS counterpart. For the current case the line element of the corresponding spacetime is given by
\begin{equation}\label{Bo.of.flat}
 d\tilde s^2=-e^{4\beta(u,\phi)}du^2+G^2d\phi^2,
 \end{equation}
which is not flat and its Ricci scalar is the same as \eqref{ricci.ads}. It is clear that the standard definition of conformal infinity does not yield   \eqref{Bo.of.flat}. However, it is possible to find \eqref{Bo.of.flat} by an anisotropic scaling of the asymptotically flat metric. This method for  definition of the conformal infinity is similar to  \cite{Horava:2009vy} where the anisotropic scaling is proposed  for definition of the conformal boundary of the  non-relativistic theories. 

Using \eqref{EM.tensor.ads} and \eqref{def.of.flat EM} we have
\begin{align}\label{EM.tensor.flat}
\nonumber \tilde T_{uu}&=\frac{e^{4\beta}\left(\partial_\phi^2\beta+(\partial_\phi\beta)^2\right)}{4\pi G^2}+\frac{{\mathcal{M}}}{16\pi G^2},\\
\nonumber \tilde T_{u\phi}&={\mathcal{N}\over 8\pi G^2},\\
\tilde T_{\phi\phi}&=\frac{\mathcal{M}e^{-4\beta}}{16\pi }-\frac{ (\partial_\phi\beta)^2 }{4\pi }.
\end{align}

Finally we find
\begin{equation}\label{Trace.anomaly.flat}
\tilde T=\tilde g^{ij}\tilde T_{ij}={1\over 4\pi}c_M\tilde R,
\end{equation}
where $\tilde R$ is the Ricci scalar of \eqref{Bo.of.flat} .This is one of the main results of the current paper\footnote{We should note that taking flat limit from the stress tensor of asymptotically AdS spacetimes  has been also studied in paper \cite{Costa:2013vza} which does not consider ultra-relativistic field theory as the dual of asymptotically flat spacetimes and our final result \eqref{Trace.anomaly.flat} is absent in that paper. }.

\section{Ultra-relativistic c-theorem }
In order to propose a holographic $C$-\hspace{.5mm}function for ultra-relativistic CFT, we first redo the holographic calculation for the asymptotically AdS spacetimes in the BMS gauge and then take the flat-space limit.
The holographic proof of c-theorem in Fefferman-Graham gauge is provided in \cite{Freedman:1999gp, Myers:2010tj}. Note that the Fefferman-Graham gauge is not appropriate for taking the flat-space limit. Therefore in this section we firstly rederive that analysis in the BMS gauge which will be appropriate for taking the flat-space limit.

To see an RG flow in boundary field theory, we should add matter to the cosmological Einstein-Hilbert theory  
\bea\label{S}
 S = {1\over 16\pi G}\int\, d^3x \sqrt{-g}\left(R+{2\over\ell^2}\right)+ S_{matter},
\eea
where "matter" could be for example a scalar  field. Note that without "matter" (\ref{S}) admits the below radial solution (a solution which depends just on "r")
\bea\label{ADS}
ds^{2} = -2dudr + r^{2}(-\frac{1}{l^{2}}d^{2}u+d^{2}x).
\eea
We demand that the boundary theory has 2d-Poincar´e symmetry. The boundary theory lives on the hypersurface at  constant "r". The most general possible 3d-bulk radial-metric in the BMS gauge which is consistent with requested symmetry can be written as
\bea\label{RGgeometry}
ds^{2} = -2dudr + e^{2B(r)}(-\frac{1}{l^{2}}d^{2}u+d^{2}x).
\eea
We propose the gravitational relativistic $C$-\hspace{.5mm}function in BMS gauge as
\bea\label{CRG}
C(r) = c~\frac{e^{-B(r)}}{B^{\prime}(r)} ,
\eea
which for that 
\bea\label{Cprime}
C^{\prime}(r) = -c~\frac{e^{-B(r)}(B^{\prime\prime}(r)+B^{\prime}(r)^{2})}{B^{\prime}(r)^{2}}.
\eea
Note that for the AdS$_{3}$ case ($B(r) = ln(r)$), the function $C(r) =c$ and $C^{\prime}(r) =0$.
The "c" is a constant but at fixed points ($AdS$ solutions) it reduces to the
central charge.

Moreover, according to the null-energy condition,  for any null vector $\xi^{\mu}$ we must have
\bea\label{Nullenergycondition}
T^{m}_{\mu\nu}\xi^{\mu}\xi^{\nu} \geq 0,
\eea
where $T^{m}$ is the matter stress-tensor. The matter stress-tensor for the theory (\ref{S})
in geometry (\ref{RGgeometry}) has this structure
\bea\label{Tmatter}
T^{m}_{\mu\nu} = \left(
  \begin{array}{ccc}
    EQ_{rr} &  EQ_{ru} & 0 \\
    EQ_{ru} &  EQ_{uu} & 0 \\
    0 & 0 &  EQ_{\phi\phi} \\
  \end{array}
\right) ,
\eea 
where
\bea
&& EQ_{rr} = -B^{\prime\prime}(r)-B^{\prime}(r)^{2},~~~~~ EQ_{uu} = \frac{e^{2B(r)}}{l^{2}}EQ_{ru},~~~~~
 EQ_{\phi\phi} = -e^{2B(r)}EQ_{ru},\cr\nonumber &&~~~~~~~~~~~~~~~~EQ_{ru} =
 -\frac{1}{l^{2}}\left(e^{2B(r)}B^{\prime\prime}(r)+2e^{2B(r)}B^{\prime}(r)^{2}-1\right).
\eea
Furthermore, by considering the vector $\xi^{\mu}$ as
\bea
\xi_{\mu} =\big (\xi_{1}(r,u,x),\xi_{2}(r,u,x),\xi_{3}(r,u,x)\big),
\eea
and solving $g^{\mu\nu}\xi_{\mu}\xi_{\nu} =0$,  it is easy to see that the general null vector in geometry (\ref{RGgeometry}) has the form:
\bea\label{Vxi}
\xi^{\mu} =\left( \frac{\sqrt{\xi_{2}^{2}l^{2}-\xi_{3}^{2}}}{l},-\frac{(\xi_{2}l+\sqrt{\xi_{2}^{2}l^{2}-\xi_{3}^{2}})l}{e^{2B(r)}},\frac{\xi_{3}}{e^{2B(r)}} \right).
\eea
Substitute the above null vector in (\ref{Nullenergycondition}) and using (\ref{Tmatter}) gives
\bea
-\frac{(\xi_{2}^{2}l^{2}-\xi_{3}^{2})}{l^{2}}\left(B^{\prime\prime}(r)+B^{\prime}(r)^{2}\right) \geq 0.
\eea
Note that the first parenthesis in the above equation is positive because it also appears in the square root term of (\ref{Vxi}). By using this result in (\ref{Cprime}) and noting that $c > 0$ for unitary theories, we see that $C^{\prime}(r)$ is semipositive. Therefore by increasing "r", $C(r)$ also increases. Note that for the metric (\ref{ADS}) the boundary lives at $r \rightarrow \infty$ which is the UV regime of the dual field theory\footnote{One of the Killing vectors of $AdS_{3}$ in the BMS gauge is $\xi^{\mu} =(\lambda r,-\lambda u,-\lambda x)$. 
Therefore by moving to the boundary, the norm of the spatial coordinate (norm of the difference between zero and arbitrary point in the boundary) will decrease. This means moving to the UV regime of the dual field theory.}. Thus by moving from  the boundary to the bulk in the gravity side
($r =\infty$ to $ r =0$) which is equivalent  to moving from the UV to the IR regime in the dual field theory, $C(r)$ has a monotonic behavior. This completes the proof of  the monotonic behavior of the gravitational relativistic $C$-\hspace{.5mm}function (\ref{CRG}) in the BMS gauge.

Let us see what happens in the flat space limit ($l \rightarrow \infty$). Firstly (\ref{RGgeometry}) in the flat space limit becomes
\bea\label{FlatRGgeometry}
ds^{2} = -2dudr +e^{2B(r)}d^{2}x,
\eea 
and (\ref{Tmatter}) in the same limit becomes
\bea\label{FlatTmatter}
T^{m}_{\mu\nu} = \left(
  \begin{array}{ccc}
    EQ_{rr} &  0 & 0 \\
   0 &  0 & 0 \\
    0 & 0 &  0 \\
  \end{array}
\right) ,
\eea 
with
\bea
EQ_{rr} = -B^{\prime\prime}(r)-B^{\prime}(r)^{2}.
\eea
Moreover the general null vector of the metric (\ref{FlatRGgeometry}) is
\bea\label{FlatVxi}
\xi^{\mu} = \bigg( -\xi_{2}, -\frac{1}{2}
\frac{\xi_{3}^{2}}{e^{2B(r)}\xi_{2}},\frac{\xi_{3}}{e^{2B(r)}} \bigg).
\eea
Substituting (\ref{FlatVxi}) and (\ref{FlatTmatter}) in (\ref{Nullenergycondition}) results in
\bea\label{FlatNullenergycondition}
-\xi_{2}^{2}\left(B^{\prime\prime}(r)+B^{\prime}(r)^{2}\right) \geq 0.
\eea 
Furthermore, the flat space limit of (\ref{CRG}) and (\ref{Cprime}) are
\bea\label{FlatC}
C(r) = 6c_{M}\frac{e^{-B(r)}}{B^{\prime}(r)},
\eea 
and
\bea\label{FlatRGflow}
C^{\prime}(r) =- 6 c_{M}~\frac{e^{-B(r)}(B^{\prime\prime}(r)+B^{\prime}(r)^{2})}{B^{\prime}(r)^{2}},
\eea
respectively. By using (\ref{FlatNullenergycondition}) and assuming $c_{M} > 0$ because of unitary reason
\cite{Bagchi:2012xr}, one can see that (\ref{FlatRGflow}) is semipositive. The semipositivity of (\ref{FlatRGflow}) guarantees  that the proposed ultra-relativistic $C$-\hspace{.5mm}function (\ref{FlatC}) has a monotonic flow from the UV to the IR regime.
\section{Discussion}
In this paper we have  derived a $C$-\hspace{.5mm}function for the ultra-relativistic CFT  which decreases from the UV to the IR region. We have also derived the conformal anomaly by allowing the boundary becomes curved whilst the bulk remains flat. In this case indeed the conformal anomaly appears and the trace of the energy-momentum tensor fails to vanish. We showed that the trace anomaly is proportional
to the ultra-relativistic central charge $c_{M}$.  The reason that central charge $c_L$ does not appear in the final result is that
for the ultra-relativistic field theories which we consider in this paper,  
$c_{L}$ is actually zero. This happens because the bulk gravitational theory preserves parity. The 
non-vanishing $c_{L}$ is achievable in the gravitational theories which are not parity invariant. For those gravitational theories, exploring the effect of $c_{L}$ in the boundary conformal anomaly is interesting.

Moreover the appearance of anomaly in the trace of holographic stress tensor reflects the fact
that in the process of holographic renormalization some counterterms are added which 
break conformal symmetry. In this paper we started with the variation of the on-shell action and those  
 counterterms do not contribute and are not relevant. Note that these type of counterterms were  found in 
 the FG gauge in the renowned paper \cite{Henningson:1998gx}. 
However the metric in FG gauge is not appropriate for the flat space limit. To see which types of counterterms are necessary to do 
the full analysis of holographic renormalization in the BMS gauge will be interesting.
\subsubsection*{Acknowledgements}
We are indebted to Mahmoud Safari for his contributions in the early part of this project.  R.F would like to thank school of particles and accelerator of institute for research in fundamental sciences (IPM) for the research facilities. 

\appendix

%\section{Construction }

\end{document}